\documentstyle[psfig,12pt]{article}

\def\refe{\par\noindent\hangindent=1.5cm}
\begin{document}
\begin{center}

{\large \bf Galactic Gradients, Postbiological Evolution and the
Apparent Failure of SETI}

\vspace{0.2cm}

\large Milan M.\ \'Cirkovi\'c

\vspace{0.05cm}

{\it Astronomical Observatory Belgrade, \\
Volgina 7, 11160 Belgrade-74, Serbia and Montenegro \\
E-mail: {\tt mcirkovic@aob.aob.bg.ac.yu}}

\vspace{0.25cm}

\large Robert J. Bradbury

\vspace{0.05cm}

{\it Aeiveos Corporation, Seattle, WA 98103, USA
\\ E-mail: {\tt bradbury@aeiveos.com}
}

\end{center}
\begin{abstract}
\noindent Motivated by recent developments impacting our view of
Fermi's paradox (absence of extraterrestrials and their
manifestations from our past light cone), we suggest a
reassessment of the problem itself, as well as of strategies
employed by SETI projects so far. The need for such reevaluation
is fueled not only by the failure of searches thus far, but also
by great advances recently made in astrophysics, astrobiology,
computer science and future studies, which have remained largely
ignored in SETI practice. As an example of the new approach, we
consider the effects of the observed metallicity and temperature
gradients in the Milky Way on the spatial distribution of
hypothetical advanced extraterrestrial intelligent communities.
While, obviously, properties of such communities and their
sociological and technological preferences are entirely unknown,
we assume that (1) they operate in agreement with the known laws
of physics, and (2) that at some point they typically become
motivated by a meta-principle embodying the central role of
information-processing; a prototype of the latter is the recently
suggested Intelligence Principle of Steven J. Dick. There are
specific conclusions of practical interest to astrobiological and
SETI endeavors to be drawn from coupling of these reasonable
assumptions with the astrophysical and astrochemical structure of
the spiral disk of our Galaxy. In particular, we suggest that the
outer regions of the Galactic disk are most likely locations for
advanced SETI targets, and that sophisticated intelligent
communities will tend to migrate outward through the Galaxy as
their capacities of information-processing increase, for both
thermodynamical and astrochemical reasons. However, the outward
movement is limited by the decrease in matter density in the outer
Milky Way. This can also be regarded as a possible generalization
of the Galactic Habitable Zone, concept currently much
investigated in astrobiology.
\end{abstract}

\begin{quotation}
\noindent {\sl If you do not expect the unexpected, you will not
find it; for it is hard to be sought out and difficult.}
\end{quotation}

\hspace{3cm} Heraclitus of Ephesos, fragment B18 (cca.\ 500 BC)

\vspace{0.6cm}

\section{Introduction}

Fermi's paradox\footnote{It would be most appropriately to call it
Tsiolkovsky-Fermi-Viewing-Hart-Tipler's paradox (for much of the
history, see Brin 1983; Kuiper and Brin 1989; Webb 2002, and
references therein). We shall use the locution "Fermi's paradox"
for the sake of brevity, and with full respect for contributions
of the other important authors.} has become significantly more
serious, even disturbing, of late. This is due to several
independent lines of scientific and technological advances
occurring during the last $\sim 10$ years:
\begin{itemize}
\item Discovery of more than 150 extrasolar planets, on almost
weekly basis (for regular updates see {\tt
http://www.obspm.fr/planets}); many of them are reported to be
parts of systems with stable circumstellar habitable zones (Noble,
Musielak, and Cuntz 2002; Asghari et al.\ 2004; Beaug\'e et al.\
2005).

\item Improved understanding of the details of chemical and
dynamical structure of the Milky Way and its Galactic habitable
zone (GHZ; Gonzalez, Brownlee, and Ward 2001). In particular, this
includes important recent calculations showing that Earth-like
planets began forming more than 9 Gyr ago and their median age is
$6.4 \pm 0.7$ Gyr, significantly more than the Earth's age
(Lineweaver 2001; Lineweaver et al.\ 2004).

\item Confirmation of the {\it rapid\/} origination of life on
early Earth (e.g.\ Mojzsis et al.\ 1996); this rapidity, in turn,
offers a strong probabilistic support to the idea of many planets
in the Milky Way inhabited by at least the simplest lifeforms
(Lineweaver and Davis 2002).

\item Discovery of extremophiles and the general resistance of
simple lifeforms to much more severe environmental stresses than
it had been thought possible previously (e.g.\ Cavicchioli 2002).
These include representatives of all three great domains of
terrestrial life ({\it Bacteria}, {\it Archaea}, and {\it
Eucarya\/}), showing that the number and variety of cosmic
habitats for life are probably much larger than conventionally
imagined.

\item Our improved understanding in molecular biology and
biochemistry leading to heightened confidence in the theories of
naturalistic origin of life (Lahav, Nir, and Elitzur 2001;
Ehrenfreund et al. 2002; Bada 2004). The same can be said, to a
lesser degree, for our understanding of the origin of intelligence
and technological civilization (e.g.\ Chernavskii 2000).

\item Exponential growth of the technological civilization on
Earth, especially manifested through Moore's Law and other
advances in information technologies (Moravec 1988; Schaller 1997;
Bostrom 2000). This includes the increased confidence in
speculations about the ultimate physical limits on computation
(Lloyd 2000).

\item Improved understanding of feasibility of interstellar travel
in both classical sense (Vulpetti 1999; Andrews 2003), and in the
more efficient form of sending inscribed matter packages over
interstellar distances (Rose and Wright 2004).

\item Theoretical grounding for various
astro-engineering/macro-engineering projects (Criswell 1985;
Badescu 1995; Badescu and Cathcart 2000; Korycansky, Laughlin, and
Adams 2001; McInnes 2002) potentially detectable over interstellar
distances. Especially important in this respect is possible
combination of astro-engineering and computation projects of
advanced civilizations, like those envisaged by Sandberg (1999).
\end{itemize}
Although admittedly uneven and partially conjectural, this list of
advances and developments (entirely unknown at the time of
Tsiolkovsky's and Fermi's original remarks, and even Viewing's,
Hart's and Tipler's subsequent re-issues) testifies that Fermi's
paradox is not only still with us more than half a century later,
but that it is more puzzling and disturbing than
ever.\footnote{One is tempted to add another item of a completely
different sort to the list: the empirical fact that we have
survived about 60 years since invention of the first true weapons
of mass destruction gives us at least a vague Bayesian argument
countering the ideas---prevailing at the time of Fermi and his
original lunch-time question---that technological civilizations
tend to destroy themselves as soon as they discover nuclear power.
This is not to contest that the bigger of part of the road toward
safety for humankind is still in front of us.} In addition, we
have witnessed substantial research leading to a decrease in
confidence in the so-called Carter's (1983) "anthropic" argument,
the other mainstay of SETI scepticism (Wilson 1994; Livio 1999;
\'Cirkovi\'c and Dragi\'cevi\'c 2005, preprint). All this is
accompanied by increased public interest in astrobiology and
related issues (e.g.\ Ward and Brownlee 2000, 2002; Webb 2002;
Cohen and Stewart 2003; Dick 2003). The list above shows,
parenthetically, that quite widespread (especially in popular
press) notion that there is nothing new or interesting happening
in SETI studies is deeply wrong.

Faced with aggravated situation vis-\`{a}-vis Fermi's paradox, the
solution is usually sought in either (i) some version of the "rare
Earth" hypothesis (i.e., the picture which emphasizes inherent
uniqueness of our planet, and hence uniqueness of human
intelligence and technological civilization in the Galactic
context), or (ii) "neo-catastrophic" explanations (ranging from
the classical "mandatory self-destruction" explanation, championed
for instance by von Hoerner or Shklovsky, to the modern emphasis
on mass extinctions in the history of life and the role of
catastrophic impacts, gamma-ray bursts, and similar dramatic
events). Both these broad classes of hypotheses are unsatisfactory
on several counts: for instance, "rare Earth" hypotheses reject
the usual Copernican assumption (Earth is a typical member of the
planetary set), and neo-catastrophic explanations usually fail to
pass the non-exclusivity requirement (but see \'Cirkovi\'c
2004a,b). None of these are clear, straightforward solutions. It
is quite possible that a "patchwork solution", comprised of a
combination of suggested and other solutions remains our best
option for solving this deep astrobiological problem. This
motivates the continuation of the search for plausible
explanations of Fermi's paradox.

Hereby, we would like to propose a novel solution, based on the
astrophysical properties of our Galactic environment on large
scales, as well as some economic and informational aspects of the
presumed advanced technological civilizations (henceforth ATCs).
In doing so, we will suggest a radically new perspective on the
entire SETI endeavor.

\section{Digital perspective and the postbiological universe}

In an important recent paper, the distinguished historian of
science Stephen J. Dick argued that there is a tension between
SETI, as conventionally understood, and prospects following
exponential growth of technology as perceived in recent times on
Earth (Dick 2003):
\begin{quotation}
\noindent But if there is a flaw in the logic of the Fermi paradox
and extraterrestrials {\it are\/} a natural outcome of cosmic
evolution, then cultural evolution may have resulted in a
postbiological universe in which machines are the predominant
intelligence. This is more than mere conjecture; it is a
recognition of the fact that cultural evolution -- the final
frontier of the Drake Equation -- needs to be taken into account
no less than the astronomical and biological components of cosmic
evolution. [emphasis in the original]
\end{quotation}
It is easy to understand the necessity of redefining SETI studies
in general and our view of Fermi's paradox in particular in this
context: for example, postbiological evolution makes those
behavioral and social traits like territoriality or expansion
drive (to fill the available ecological niche) which are---more or
less successfully---"derived from nature" lose their relevance.
Other important guidelines must be derived which will encompass
the vast realm of possibilities stemming from the concept of
postbiological evolution. In particular, we follow the {\it
Intelligence Principle\/} of Dick (2003), stating that
\begin{quotation}
\noindent In sorting priorities, I adopt what I term the central
principle of cultural evolution, which I refer to as the
Intelligence Principle: {\it the maintenance, improvement and
perpetuation of knowledge and intelligence is the central driving
force of cultural evolution, and that to the extent intelligence
can be improved, it will be improved.} [emphasis in the original]
\end{quotation}
Before we explore the logical consequences of the Intelligence
Principle for SETI further, let us emphasize that the study of
Dick (2003) is not an isolated instance. Very similar thinking is
clearly emerging in various other fields and related to a plethora
of different problems. Considerations of postbiological evolution
may be fruitfully related to the {\it megatrajectory\/} concept of
Knoll and Bambach (2000), who cogently argue that astrobiology is
the ultimate field for verification or rejection of our biological
concepts. In relation to the old problem of progress (or its
absence) in the evolution of life on Earth, Knoll and Bambach
offer a middle road encompassing both contingent and convergent
features of biological evolution through the idea of a
megatrajectory:
\begin{quotation}
\noindent We believe that six broad megatrajectories capture the
essence of vectorial change in the history of life. The
megatrajectories for a logical sequence dictated by the necessity
for complexity level $N$ to exist before $N+1$ can evolve... In
the view offered here, each megatrajectory adds new and
qualitatively distinct dimensions to the way life utilizes
ecospace.
\end{quotation}
The six megatrajectories outlined by the biological evolution on
Earth so far are: (i) from the origin of life to the "Last Common
Ancestor"; (ii) prokaryote diversification; (iii) unicellular
eukaryote diversification; (iv) multicellularity; (v) invasion of
the land; and (vi) appearance of intelligence and technology. {\it
Postbiological evolution may present the seventh megatrajectory,}
triggered by the emergence of artificial intelligence at least
equivalent to the biologically-evolved one, as well as the
invention of several key technologies of roughly similar level of
complexity and environmental impact, like molecular nanoassembly
(Phoenix and Drexler 2004) or stellar uplifting (Criswell 1985).
ATCs can be regarded as instantiations of this seventh (or
possible higher) megatrajectory. It is not necessary to assume
that the seventh megatrajectory represents the complete or partial
abandonment of the biological material substratum of previous
evolution, although that is certainly one of the options. Rather,
the mode of evolution is likely to change from the Darwinian one
dominating previous six megatrajectories, to a sort of
aggregative, intentional, quasi-Lamarckian mode characteristic for
highly developed cultural entities. We shall repeatedly return to
this important point, which in a sense obviates further rather
superficial speculation about the {\it detailed\/} structure of
ATCs.

A natural extension of the Intelligence Principle is what can be
called the {\it digital perspective\/} on astrobiology: after a
particular threshold astrobiological complexity is reached, the
relevant relations between existent entities are characterized by
requirements of computation and information processing. It is
related to the emergent computational concepts not only in
biology, but in fundamental physics, cosmology, social sciences,
etc. One particular consequence of the digital perspective,
dealing with the thermodynamics of computation, we shall now
argue, will allow us a glimpse of a novel view of the generic
evolution of the intelligent communities in the Galactic context,
including a new solution of the old Fermi's puzzle. The digital
perspective also indicates that we should either entirely abandon
or significantly modify Kardashev's (1964) classification of
extraterrestrial intelligent communities, one of the mainstays of
classical SETI studies.

What limits prospects of postbiological evolution guided by the
Intelligence Principle? In order to answer this question, we need
to consider limitations imposed by physics on the classical theory
of computation. As almost anybody having practical experience with
computers will have experienced, heat is an enemy of computation.
In contrast to other obstacles and difficulties facing highly
imperfect computers of today (like limited storage space, dust
gathering on chips, or inefficiency of their human operators), the
problem of heat dissipation is a consequence of the laws of
physics. Therefore, we conjecture that  this problem will remain
{\it the\/} enemy of efficient computation for advanced technical
civilizations, and that it will have a dominant effect on
policy-making of such advanced societies.

Thermodynamics of computation has, historically, been motivated by
Maxwell's demon "paradox" which led to great breakthroughs of
Szilard, Brillouin, and Landauer. One of its most important
results, often called Brillouin inequality is the fundamental
property of the information content available for processing in
any sort of physical system (Landauer 1961; Brillouin 1962):
\begin{equation}
\label{bril1} I \leq I_{\rm max},
\end{equation}
where the limiting amount of information $I_{\rm max}$ (in bits)
processed using energy $\Delta E$ (in ergs) on the processor
temperature $T$ (in K) is given as
\begin{equation}
\label{bril2} I_{\rm max} = \frac{\Delta E}{k_B T \ln 2}= 1.05
\times 10^{16} \frac{\Delta E}{T}.
\end{equation}
Here, $k_B$ is the Boltzmann constant. Obviously, computation
becomes more efficient as the temperature of the heat reservoir in
contact with the computer is lower. In the ideal case, no energy
should be expended on cooling the computer itself, since that
expense should be added to the energy cost of logical steps
minimized by (\ref{bril2}). The most efficient heat reservoir is
the universe itself, which {\it far from local energy sources\/}
like stars and galaxies, has the temperature of the cosmic
microwave background (henceforth CMB; Wright et al.\ 1994)
\begin{equation}
\label{ccmb}
T_{\rm CMB} = 2.736 \pm 0.017 \; {\rm K}.
\end{equation}
However, this is an ideal case, since ATCs cannot have their
computers in thermal equilibrium with CMB for astrophysical
reasons. In the rest of this paper we shall investigate how close
approach to this ideal case is feasible.

It has already been repeatedly suggested that our descendants, in
particular if they cease to be organic-based, may prefer
low-temperature, volatile-rich outer reaches of the Solar System.
Thus, they could create what could be dubbed "circumstellar
technological zone" as different and complementary to the famous
(and controversial) "circumstellar habitable zone" in which life
is, according to most contemporary astrobiological views, bound to
emerge. We propose to generalize this concept to the Galaxy (and
other spiral galaxies) in complete analogy to GHZ (Gonzalez et
al.\ 2001; Lineweaver et al.\ 2004). It is not necessary, or
indeed desirable, for our further considerations to make the
notion of ATCs more precise. The diversity of postbiological
evolution is likely to at least match, and probably dwarf, the
diversity of its biological precedent. It is one particular
feature---information processing---we assume common for the
"mainstream" ATCs. Whether real ATCs can most adequately be
described as "being computers" or "having computers" is not of key
importance for our analysis; we just suppose that in either case
the desire for optimization of computations will be one of
important, if not the most important, desires of such advanced
entities. It is already clear, from the obviously short and
limited human astronautical experience, that postbiological
evolution offers significant advantages in this field (Parkinson
2005).

\section{Galactic temperature gradient}

The famous article by Freeman Dyson (1960) proposing search for
large-scale engineering projects (like eponymous Dyson shells) as
the signposts of the presence of advanced extraterrestrial
intelligence provoked much discussion henceforth. One very
important contribution was the early suggestion of the
distinguished computer scientist and AI pioneer Marvin Minsky
(1973) in a debate following Dyson's talk at the celebrated
Byurakan conference in 1971, that advanced computers would utilize
the temperature of cosmic microwave background as a heat sink
(\ref{ccmb}).\footnote{Parenthetically, in the same debate Minsky
presciently suggested infeasibility of conventional SETI due to
the impossibility of distinguishing the signal from the Gaussian
noise (cf.\ Lachmann, Newman, and Moore 2004).} This particular
idea is wrong in the specifics, at least for the younger and most
accessible ATCs, but it gives us an important hint as to what
should we be searching for. Subsequently, other astro-engineering
projects---sometimes called megaprojects or macroprojects---aimed
at optimization of resources at ATCs' disposal have been proposed,
notably Jupiter Brains (Sandberg 1999; for early history see
Bradbury 1997; Perry E. Metzger, private communication to RJB May
20, 1998) and Matrioshka Brains (Bradbury 2001).

What is the temperature of a solid body (like a Dyson shell, a
Matrioshka brain, or a Jupiter brain\footnote{For the purpose of
the present discussion, we use the placeholder "solid body" for
any macroscopic body not made of gas or liquid. Thus, sizes of
solid bodies we consider range roughly from $10^{-5}$ cm (an
interstellar dust grain) to $10^{13}$ cm (a Dyson shell).} in
thermal equilibrium with the surrounding interstellar space? The
dominant factor is the spatial distribution of the interstellar
radiation field (henceforth ISRF), especially at short wavelengths
(optical and UV). It can be shown that the most important way of
energy transfer to a solid body in by far the largest fraction of
the Galactic interstellar medium (ISM), is the absorption of
photons while collisions with atoms and ions are unimportant. For
example, ambiental ultraviolet flux close to the Solar circle of
about $10^{10}$ photons m$^{-2}$ s$^{-1}$ nm$^{-1}$ will deposit
about a $10^{-20}$ J s$^{-1}$ to a dust grain (with unity
absorption factor, for simplicity), while collisions deposit
$\simeq 10^{-26} \, n_{\rm ISM}$ J s$^{-1}$, where $n_{\rm ISM}$
is the ISM number density in cm$^{-3}$. Since on the average
$\langle n_{\rm ISM} \rangle = 1$ cm$^{-3}$, we perceive how
unimportant collisions which form our laboratory definition of the
"thermal equilibrium" are in the interstellar space.

Neglecting collisions, the temperature will be given by solving
the radiative equilibrium equation (e.g.\ Dyson and Williams 1980)
\begin{equation}
\label{osnov} \int F(\lambda) Q_{\rm abs} (a, \lambda) \, d
\lambda = \int Q_{\rm abs}(a, \lambda) B(\lambda, T) \, d \lambda,
\end{equation}
where $F(\lambda)$ is the energy flux of ISRF, $Q_{\rm abs}$ is
the absorption coefficient, and the Planck black-body function is
given as
\begin{equation}
\label{plank} B(\lambda, T) = \frac{2hc}{\lambda^3}
\frac{n_\lambda^2}{\exp \left( \frac{hc}{kT \lambda} \right) -1}
\;.
\end{equation}
ISRF is created mainly by massive stars of Population I,
concentrated in the Galactic disk. Typical value of the energy
density of ISRF in vicinity of the Sun is $U = 7 \times 10^{-13}$
ergs cm$^{-3}$, which does not include the CMB contribution, which
has $U_{\rm CMB} = 4 \times 10^{-13}$ ergs cm$^{-3}$. We use the
conventional assumption of the exponential disk (e.g.\ Binney and
Merrifield 1998) with the luminosity density approximated by
\begin{equation}
\label{distrib} j(R,z) = \frac{I_0}{2 z_0}\exp \left( -
\frac{R}{R_d} - \frac{|z|}{z_0} \right),
\end{equation}
where $I_0$ is the central surface brightness, $z_0$ is the
scale-height, and $R_d \approx 3$ kpc is the disk scalelength.
From this, we obtain the disk surface brightness as
\begin{equation}
\label{sdistrib} I(R) = \int\limits_{-\infty}^{+\infty} j(R, z) \,
dz = I_0 \exp \left( - \frac{R}{R_d} \right),
\end{equation}
which agrees with observations in other disk galaxies. It seems
clear that ISRF will decline with galactocentric distance, and
thus the equilibrium temperature will decline too, enabling more
and more efficient computation, as per (\ref{bril2}). No detailed
studies of the radiation field temperature distribution in the
Milky Way disk exist so far, we suggest the following rough
estimates.

\begin{center}
\begin{figure}
\psfig{file=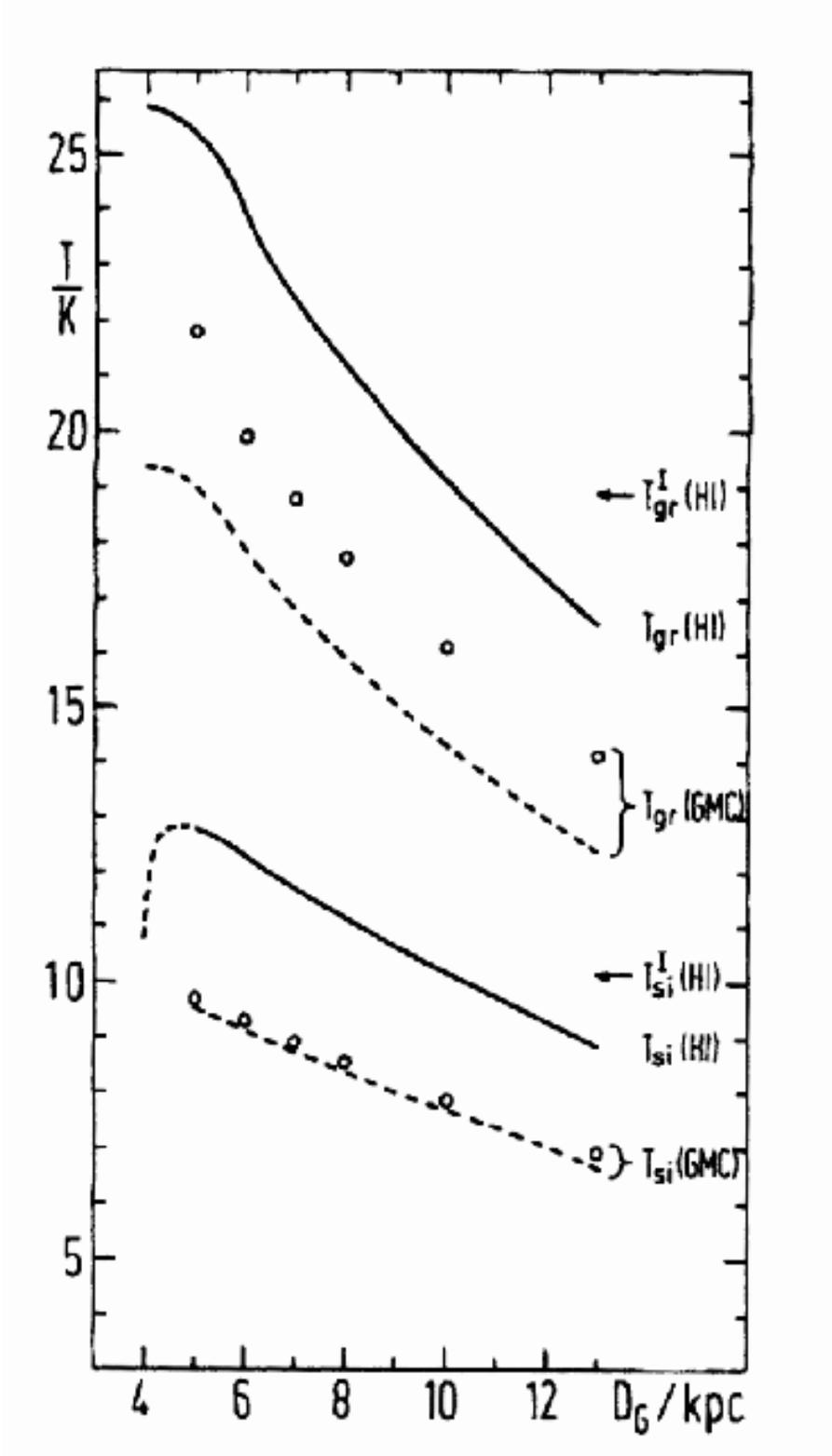,width=12cm} \caption{Temperature of dust
grains in equilibrium with ISRF for various galactocentric
distances given by Mathis et al.\ (1983); courtesy of Prof. John
S.\ Mathis. \label{pic1}}
\end{figure}
\end{center}

\begin{figure}
\psfig{file=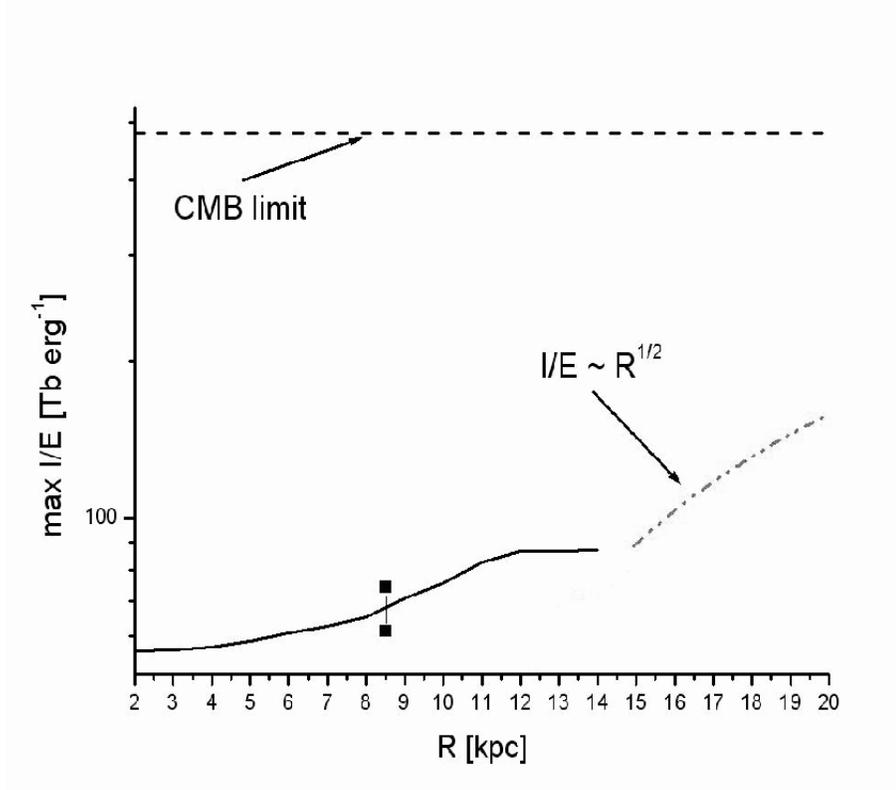,width=12cm} \caption{Maximal quantity of
information per unit expended energy processed by a computer in
equilibrium with the radiation field at various galactocentric
distances in the Milky Way (with the location of the Solar System
indicated by a vertical bar). The limit due to the cosmic
microwave background is also given. We see that the efficiency
limit rises considerably for $R > 10$ kpc. At large galactocentric
distances, a simple scaling argument indicates that the efficiency
will tend to rise $\propto \sqrt{R}$ until it reaches the CMB
limit in the "true" intergalactic space; however, those external
regions are almost devoid of baryonic matter, and even the density
of cold dark matter (CDM) particles becomes exceedingly small.
\label{pic2}}
\end{figure}

Part of the answer can be gauged by comparison with the existent
natural solid bodies in such thermal equilibrium, namely
interstellar dust grains. In Figure \ref{pic1}, we see results of
the detailed models of the Galactic temperature distribution of
dust grains (Mathis, Mezger, and Panagia 1983; Cox, Kr\"{u}gel,
and Mezger 1986). Using the same values of ISRF and correcting for
the emission efficiency of larger solid objects we get the results
presented in Figure \ref{pic2}, for computation efficiency defined
as the maximal number of bytes processed per erg of the invested
energy in terabytes ($10^{12}$ bytes) per erg.

In reality, we need to take into account the inhomogeneities in
the interstellar medium, especially giant molecular clouds. The
interiors of molecular clouds are impenetrable to short wavelength
radiation, and present some of the coldest locales in the Milky
Way ($T \sim 10$ K). However, the interiors of giant molecular
clouds are also sites of vigorous massive star formation, so these
low-temperature locales are quite irregular and transient
phenomena, assembling and dissembling on timescales of $\sim 10^6$
yrs, which is probably unacceptable from the point of view of most
ATCs (which we suppose stable at longer timescales by definition).

There have not been any studies of the ISRF for distances larger
than about 14 kpc (Prof.\ John S.\ Mathis, private communication).
At some point for large galactocentric distances (larger than the
Holmberg radius $\sim 20$ kpc), practically all sources of ISRF
are located at smaller $R$, so we can use the simplest
approximation of galaxy as a point source. If with $T_D (R)$ we
denote the temperature of a large solid object (a Dyson shell,
say) at galactocentric distance $R$, a simple scaling relationship
holds:
\begin{equation}
\label{propt} \sigma T_D^4 (R) \propto L_\ast R^{-2},
\end{equation}
where $\sigma$ is Steffan-Boltzmann constant, $L_\ast \approx 4.9
\times 10^{10} \; L_\odot$ the Galactic luminosity in the
absorbing bands. Taking into account both (\ref{bril2}) and
(\ref{propt}), we obtain the general scaling relation in the
outermost regions:
\begin{equation}
\label{propt2} \left( \frac{I}{E} \right)_{\rm max} \propto
\sqrt{R}.
\end{equation}
At large distances, the CMB limit will be reached asymptotically.
Other issues to be taken into account in a future more complete
treatment of the problem of habitability of the Galaxy from the
point of view of (probably postbiological) ATCs are the following:
\begin{itemize}
\item Cosmic ray heating, which is important even in the interiors
of the densest molecular clouds (it dominates heating and
ionization mechanisms there and initiates all chemical reactions
in cold environments); the cosmic ray energy density at the Solar
circle is about $U_{cr} = 2.4 \times 10^{-12}$ erg cm$^{-3}$
(Webber 1987), but falls off in a complicated manner with
galactocentric distance (including Parker instability, etc.).

\item Supernovae, especially of the core-collapse Type II and Type
Ic, which tend to be concentrated in spiral arms and other regions
of intense star-formation (for the astrobiological significance of
supernovae for planetary biospheres, see e.g. Tucker and Terry
1968; Ruderman 1974; Hunt 1978; Collar 1996).

\item Much rarer and more dramatic events, Galactic gamma-ray
bursts (the longer ones associated with hypernovae and perhaps
also shorter ones caused by neutron stars' mergers), capable of
adversely influencing planetary biospheres over a large part of
the Galaxy (e.g.\ Thorsett 1995; Scalo and Wheeler 2002; Dar and
De R\'{u}jula 2002; Melott et al.\ 2004; Thomas et al. 2005).
Possible Galactic nuclear activity falls in the same category (see
below).

\item Other thermodynamical issues related to computational needs,
notably bit-erasure costs, as well as bandwidth and latency issues
(Sandberg 2000).
\end{itemize}
It is significant that both radiative and kinetic energy inputs
from supernovae and related events are adverse to the computation
efficiency of ATCs. All these effects are falling off with the
galactocentric distance, and become very small for $R
> 15$ kpc. In the inner parts of the Galaxy, the same factors
which preclude habitability (mainly supernovae and gamma-ray
bursts) act to preclude computation as well. In addition, the
issue of the nuclear activity of the Milky Way and spiral galaxies
in general, may be important for astrobiological evolution of
those regions. It has been proposed by Clarke (1981) in an
interesting early paper, as a mechanism of global regulation
preventing life and intelligence from arising in the entire
Galaxy; see also Clube (1978) and LaViolette (1987). Although it
seems now that the original idea is implausible in light of the
specific conditions in Milky Way's nucleus, we should still be
cautious, since the recent research unveiled tremendous nuclear
outbursts in distant objects (e.g.\ McNamara et al.\ 2005).

There is another large-scale gradient in the Milky Way recently
firmly established by observations: the metallicity gradient.
Average metal abundances of stars and ISM are rather well
described by the radial gradient of $\nabla Z \simeq 0.07$ dex/kpc
(Hou, Prantzos, and Boissier 2000; Tadross 2003). It could seem at
first that this is adverse for ATCs in the outer regions of the
Galaxy, thus counteracting the trend of increasing computing
efficiency described above. But, the Intelligence Principle
suggests something different, when we take into account that the
chemical enrichment causing the gradient is entirely the product
of stellar nucleosynthesis; primordial composition was uniformly
metal-free. Stellar nucleosynthesis is appallingly inefficient
process by ATC standards; it converts $\sim 1$\% of the rest mass
to energy and most of that created energy left the Galaxy a long
time before the emergence of first ATCs. Thus, baryonic matter in
the primordial composition would, in principle, be desirable for
advanced optimization of computation. ATCs would subsequently be
able to create heavier nuclei by controlled nuclear fusion and
minimize the energy leak per unit of created entropy.

\section{Migration hypothesis}

Taking all this into account, we suggest the "migration
hypothesis": ATCs will tend to move their computing facilities
toward the colder regions of the Milky Way in order to make their
information processing as efficient as possible. In general case,
this would mean the outskirts of the Galaxy, but the interiors of
the giant molecular clouds could also serve as local foci for the
advanced information processing.\footnote{Insofar as the other
risks, from the point of view of ATCs and the Intelligence
Principle, mentioned above can be avoided: notably star-formation
bursts and associated Type II supernovae.} If the postbiological
evolution is predominant, as suggested by Dick (2003) and other
recent authors, this would mean that the entire ATC will tend to
migrate outward from its original location in the GHZ toward a
convenient location in the Galactic "technological zone" (GTZ)
with temperature low enough to increase computing efficiency.
Although such a migration will seem expensive at first glance, it
is not necessarily so: postbiological civilizations are likely to
be small, compact, stable over astrophysical timescales and would
be able to travel as redundant information storages at small
speeds with negligible energy expenditures. Almost all energy will
be needed for acceleration and deceleration. Starting at a
particular galactocentric distance, it is not difficult to
calculate that any specific transportation cost will be covered by
increased computation efficiency on timescales short compared to
the astrophysical timescales (stellar Main Sequence lifetimes; cf.
Zuckerman 1985) or even the timescale of the travel itself!
Interestingly enough, bold suggestions for the possible
technologies of such interstellar migrations already exist in the
literature, notably in the form of "stellar engines" (e.g.\
Badescu and Cathcart 2000) or gravitational assists (Prado 1996;
Vulpetti 1999).

What limits the outward migration of ATCs? This is largely a
context-dependent issue, but the most plausible limit is set
simply by available supply of matter. Depending on whether ATCs
get to use non-baryonic dark matter, whose density roughly varies
in accordance with the isothermal profile ($\rho \propto R^{-2}$),
or only baryonic matter, which falls off exponentially
(\ref{distrib}), the maximal distance a cost-mindful ATC may be
located at will greatly vary. But in each case, it is a {\it
well-defined value}, which limits the Galactic technological zone
from the outside. The concept of GTZ should not be understood as
strict and immutable: it just indicates higher relative density of
"technologized" matter than elsewhere. ATCs can arise and function
elsewhere (in the same sense as life can arise outside of the
classical circumstellar habitable zone; e.g.\ in Europa-like
subglacial oceans), but the probability of finding them is not
uniform; on the present hypothesis, the maximum probability will
be located in the ring on the periphery of the Milky Way.

It is important to understand that while we do not doubt that ATCs
will eventually have astro-engineering means to prevent any
individual catastrophic occurrences like supernovae or
GRBs,\footnote{This could be achieved through technologies already
envisaged, like stellar lifting and long-term orbit
modifications.} we doubt that it can ever become worthwhile to
manage and police the Galaxy in this manner. Energy/information
and time cost are likely to remain too high in all epochs. On the
contrary, it seems probable that any rational cost-benefit
analysis would favor migration to the Galactic rim rather than the
costly and risky "policing" strategy.

Migration in physical space will be analogous to the prior
migration of the bulk of civilization's interests and pursuits
from physical to digital space. This presents an additional factor
helping explain Fermi's paradox: advanced civilizations based on
an optimized computronium (i.e.\ whatever substrate upon which
computation can take place) infrastructure have little need for
conversations with human-level individuals or even civilizations
whose thought capacities are trillions of times less than their
own (cf. \'Cirkovi\'c and Radujkov 2001). In contrast, they may
have an interest in leaving our civilization and other "late
comers" to their own unique development path so as to increase the
potential diversity and information content of the Galaxy. As
elaborated by Bradbury (2001), this is due to the large phase
space of what can be constructed using molecular nanotechnology
and the difficulties in proving that the computational
architectures previously adopted to support advanced civilizations
are, in fact, "optimal". ATCs may need less developed
civilizations for the "dumb luck" they may have in developing an
unexplored quadrant of the phase space of what may be designed and
assembled in support of the evolution of intelligence. An
additional consequence of the outward migration would be a
dramatic increase in the average distance between ATCs; this
circumstance will be important for assessing practical SETI
prospects below.

Compared with the usual SETI assumptions, the present hypothesis
favors a small, compact, highly efficient ATCs, in both biological
and postbiological cases. In contrast to the usually assumed model
of expanding "colonial empire" from human history (which confronts
us with the gravest form of Fermi's paradox), the present picture
would rather use a model of a "city-state"---if anything from the
human history is to be even remotely analogous to the generic
pathway of ATCs, which is itself a doubtful proposition. It is too
often forgotten (both among SETI proponents, as well as the
contact pessimists) that colonial expansion has been an exception,
rather than the rule in human history so far; our Western-centric
attitude should not blind us into accepting a wrong model for
civilizational behavior. Countless city-states, be they in ancient
Greece, pre-Aryan India, Babylonia, medieval Italy, Germany or
Russia, pre-Incan Andes or Mayan Mexico, have all together much
longer and stronger traditions than imperial powers, of which
there are no more than two dozen examples altogether, from Assyria
to the USA. Even in the cases where cities and other smaller
organizational units have been peacefully or otherwise
incorporated into a larger whole, this was often regarded as
optimization of resources and management, and clear limits to
growth have been set in advance; examples in this respect range
from Achaean League, to Hansa, to Swiss Confederation, to China
after Ch'in unification. It is exactly this understanding of
limits (or resources and communication) which made the longevity
of civilizations like the Chinese, or organizations like the Roman
Catholic Church so prominent in the human history so far.
\textit{Vice versa}, it was disregard for these limits which
contributed to downfalls of all historical empires.

As noticed by Gould (1989), the normative concept of "progress
through conquest and displacement" is intimately linked with the
Victorian "chain of being" fallacy. According to this view, all
lifeforms have their exact position in the chain ranging from the
most primitive {\it Archaea\/} to the gentlemen with white hats
doing a noble job of conquering savages all around the world. This
view has been abandoned in practically all fields---except,
ironically enough, SETI studies. In general, SETI is mostly in the
same shape and with the same set of philosophical, methodological
and technological guidelines as it was in the time of its pioneers
(Drake, Sagan, Shklovskii, Bracewell, Oliver, Morrison) in 1960s
and 1970s.\footnote{For a prototype "Galactic Club"
optimistic---or naive---view of that epoch, exactly 30 years old,
see Bracewell (1975). Early SETI literature abounds in such
enthusiasm.} In contrast, our views of astrophysics, biology, and
especially, computer science---arguably the three key scientific
disciplines for SETI---changed revolutionarily, to put it mildly,
since that epoch. The present study is an attempt to break this
mold and point serious modern alternatives to the old-fashioned
SETI philosophy.

The present approach is similar to the one favored by Dyson
(2003), who suggests searching for life at distant objects of the
Solar System (and other planetary systems). Although Dyson only
considers primitive life, this can be easily generalized to life
of higher level of complexity and even
intelligence.\footnote{Parenthetically, the same line of reasoning
suggests that the search for extraterrestrial artifacts (SETA)
should concentrate to the outskirts of the Solar System, notably
the Kuiper belt objects and even Oort cloud comets. Low-profile
digital approach would warrant maximization of the information
processing for the hypothetical ATC probes also.} In his other
writings, Dyson has suggested supercivilizations of various sorts
whose activities can be detected even if they are not actively
engaged in an effort to contact and communicate with other
societies; for an early review of such ideas, see Lemarchand
(1995). In our view, the migration hypothesis can solve Fermi's
paradox, since the truly advanced societies, i.e.\ those who
survive the bottleneck presented by the threat of self-destruction
through warfare or accident will tend to be located at the
outskirts of the Milky Way, outside of the main thrust of SETI
projects so far. The very same traits making ATCs capable of
migrating and utilizing resources with high efficiency
(compactness, high integration, etc.) will tend to make them
systematically hard to detect from afar. This is in diametrical
opposition to views of many early SETI researchers---recently
brought to a sort of the logical extreme by Weinberger and Hartl
(2002)---that ATCs will indulge in extravagant spending in order
to achieve interstellar communication, even if only nominal. The
same applies {\it mutatis mutandis\/} to the large-scale
interstellar travel to diverse targets; the nature of the
postbiological megatrajectory is not likely to include any gain
from the scattering of pieces of an ATC all over the
Galaxy.\footnote{This is related to the possibility of
postbiological ATCs being what Bradbury (2001) dubs "distributed
replicated intelligence[s]". Although elaboration of this
admittedly speculative concept is far beyond the scope of the
present study, it is enough to mention the intuitively clear
picture in which, once a threshold of complexity is reached, it is
very hard to separate an intelligent part from the whole of the
ATC to any distance making the latency problems due to the finite
speed of light important. This may pose problems for anything but
the simplest, low-profile, interstellar drones or passive
inscribed matter packages.} One of the SETI pioneers, Benjamin
Zuckerman proposed in 1985 that stellar evolution is an important
motivation for civilizations to undertake interstellar migrations
(Zuckerman 1985). Although arguments presented in that study seem
outdated in many respects, it is significant that the mass
migration idea has been presented even in the context of classical
SETI studies, biological evolution and pre-digital perspective. It
seems implausible that any but the most extreme conservative
societies would opt to wait to be forced to migration by slow and
easily predictable process like their domicile star leaving the
Main Sequence.

Not surprisingly, some of the ideas presented here have been
forefathered in a loose form within SF discourse. Karl Schroeder
in "Permanence" not only formulated an unrelated answer to Fermi's
question, but, more pertinently, envisaged the entire Galaxy-wide
ecosystem based on brown dwarfs (and halo population in general)
and low-temperature environment (Schroeder 2002; see also
\'Cirkovi\'c 2005). The idea of a new megatrajectory comprising
"mainstream" evolution of ATCs and containing the theoretical
explanation of Fermi's paradox has been beautifully discussed by
Stanislaw Lem (especially 1987, but see also Lem 1984). Most
strikingly, the idea of ATCs inhabiting the outer fringes of the
Milky Way has been suggested---though without the thermodynamical
rationale---by Vernon Vinge in "A Fire upon the Deep" (Vinge
1991). Vinge vividly envisages "Zone boundaries" separating dead
and low-tech environments from the true ATCs inhabiting regions at
the boundary of the disk and high above the Galactic plane. This
is roughly analogous to the low-temperature regions we outlined as
the most probable Galactic technological zone.

\section{Discussion: failure of the conventional SETI perspective}

In building of the migrational solution to Fermi's puzzle, we have
relied on the following set of assumptions:
\begin{enumerate}
\item The Copernican principle continues to hold in astrobiology,
i.e.\ there is nothing special about the Earth and the Solar
System when considerations of life, intelligent observers or ATCs
are made.

\item Laws of physics (as applied to the classical computation
theory and astrophysics) are universally valid.

\item Naturalistic explanations for the origin of life,
intelligence and ATCs are valid.

\item The Milky Way exhibits well-established gradients of both
baryonic matter density and equilibrium radiation field
temperature.

\item Habitable planets occur naturally only within GHZ (which
evolves in a manner roughly understood), but ATCs are not in any
way limited to this region.

\item We assume \textit{local} influences both of and on ATCs.
Thus, we disregard speculative ideas about wormholes, "basement
universes", etc. Interstellar travel is feasible, but is bound to
be slow and expensive (for anything larger than nanomachines) at
all epochs.

\item Astro-engineering on the scales significantly larger than
the scale of a typical planetary system (e.g., on the parsec scale
and above) will remain difficult and expensive at all epochs and
for all ATCs.

\item ATCs will tend to maximize efficiency of
information-processing, no matter how heterogeneous their
biological, cultural, etc.~structures and evolutionary pathways
are.
\end{enumerate}
These assumptions are, of course, of varying validity and
importance. Items 1, 2, and 3 are essential methodological
guidelines of the entire scientific endeavor; although 1 has
recently become controversial within "rare Earth" theorists'
circle, there is still no compelling reasons for relinquishing it.
Assumption 4 is an empirical fact, and 5 is quite close to it.
Assumptions 6 and 7 are conservative extrapolations of our limited
scientific and technological perspective, but in our view should
be retained until the contrary positions can be verified. In
particular, absence of the Galaxy-size astro-engineering effects
in external galaxies (Annis 1999b) strongly supports assumption 7.

Most controversial, of course, is the culturological (or
meta-ethical) assumption 8. One way to justify it is to observe
the alternative long-term strategies in a given cosmological
setting. Ultimately, ATCs will face the limits of cosmology and
fundamental physics (Tipler 1986; Adams and Laughlin 1997;
\'Cirkovi\'c 2004c); their vastly improved predicting capacities
will enable them to obtain high-resolution models of such
situations far in advance. Two limits seem reasonable: evolving
into either pure pleasure seeking and hedonism in the most general
sense (a "Roman empire" analogue) or onto a pathway toward the
greatest accomplishments possible along their individual
development vector (a "Greek Olympics" analogue). In either
situation they will seek the greatest computational capacity and
efficiency possible to support these activities.

We wish to re-emphasize the absence of exotic physics or
inconceivably advanced technology in our analysis. Its central
piece, Brillouin inequality is valid for classical computation. If
much discussed (in theory) quantum computation becomes practical
possibility, it might not be bound by it (although an analogous
constraint, Margolus-Levitin bound might step in its place; cf.
Dugi\'c and \'Cirkovi\'c 2002). On the more exotic/SF side of the
story, one might imagine creating wormholes to non-local sources
of usable energy, or even entire "basement universes" envisaged by
Linde (1990, 1992; see also Garriga et al.\ 2000).

The migration hypothesis smoothly joins with the global
catastrophic solutions, such as those proposed by Clarke (1981)
and Annis (1999a; see also Norris 2000). In those scenarios, there
is a {\it global regulation mechanism\/} for preventing the
formation of complex life forms and technological societies early
in the history of the Galaxy. Such a global mechanism could have
the physical form of $\gamma$-ray bursts, if it can be shown that
they exhibit sufficient lethality to cause mass biological
extinctions over a large part of the volume of the Galactic
habitable zone (Scalo and Wheeler 2002; see also Thorsett 1995;
Melott et al.\ 2004). However, since the regulation mechanism
exhibits secular evolution, with the rate of catastrophic events
decreasing with time, at some point the astrobiological evolution
of the Galaxy will experience a change of regime. When the rate of
catastrophic events is high, there is a sort of quasi-equilibrium
state between the natural tendency of life to emerge, spread,
diversify, and complexify, and the rate of destruction and
extinctions. When the rate becomes lower than some threshold
value, intelligent and space-faring species can arise in the
interval between any two extinctions and make themselves immune
(presumably through technological means) to further
extinction/causing events.\footnote{In fact, an alternative
definition of ATCs may consist of the requirement that an
intelligent community is immune, {\it as a community}, to all
kinds of natural catastrophes.} The migration hypothesis
complements such catastrophic solutions to Fermi's puzzle, since
it adds another layer to the "Great Filter" (Hanson 1998)
explaining the absence of ATCs or their manifestations. Annis' and
related hypotheses suggest that ATCs are both rarer and younger
than we would naively expect based on uncritical gradualism; the
migration hypothesis presented here indicates that even those
which exist at present would be hard to detect due to their
peripheral distribution, as well as other difficulties related to
their postbiological evolution. In other words, not-yet-ATCs are
decimated by catastrophes, while ATCs---who are immune to such
contingencies---have predominantly migrated to the Galactic rim
long time ago.

An objection that the proposed solution violates Occam's razor
must be considered. William of Occam a 14th century English
Franciscan, strongly espoused nominalism against the Platonic
concept of ideal types as entities in a realm higher than material
existence (a viewpoint conventionally known as realism). Occam
devised his famous motto, \textit{non sunt multiplicanda entia
praeter necessitatem\/} (entities are not to be multiplied beyond
necessity), as a weapon in this philosophical battle---an argument
against the existence of an ideal Platonic realm (for nominalists
regard names of categories only as mental abstractions from
material objects, and not as descriptions of higher realities,
requiring an additional set of unobserved ideal entities, or
essences). Occam's razor, in its legitimate application, therefore
operates as a logical principle about the complexity of an
argument, not as an empirical claim that nature must be maximally
simple. It is exactly this often underappreciated point which
makes the present solution to Fermi's paradox actually simpler
than most of the alternatives. Consider, for instance, the
hypothesis that a hundred prospective (independently arising) ATCs
randomly distributed over the Milky Way disk destroyed themselves
through internal warfare before leaving their home planets. It
is---apart from the appeal to a mystical and universal
\textit{fatum}---an excessively complex hypothesis, relying on
explanation of the observed "Great Silence" through a hundred
\textit{groups\/} of both logically and spatio-temporally disjoint
causes. Contrariwise, the migration hypothesis proposed here
suggests that a fraction of these civilizations will, upon
surviving the filter of natural and artificial catastrophes,
essentially drop out of sight through optimization of computing
resources (implying preferred peripheral distribution in the
Galaxy, not wasting energy on inefficient communication, etc.).
This can be, in turn, reduced to a small number of causes,
essentially those presented as the assumptions 1$-$8  above.

Once adopted as a viable solution to Fermi's paradox, the
migration hypothesis presented here has both theoretical and
practical consequences. First of all, inconvenient location of
most of ATCs as observed from the Solar System, coupled with
realization that distinguishing signal from noise is much harder
than usually thought (Lachmann et al.\ 2004) and may even be
completely substituted by inscribed-matter messages (Rose and
Wright 2004), are sufficient to explain the lack of results in
SETI projects so far. Some of the SETI pioneers have been very
well aware of this and warned about it (notably Sagan 1975); these
cautious voices have been consistently downplayed by the SETI
community. We conclude that the conventional radio SETI assuming
beamed broadcasts from targets within Solar vicinity (e.g.\
Turnbull and Tarter 2003) is ill-founded and has little chance of
success on the present hypothesis. {\it It is a clear and testable
prediction of the present hypothesis that the ongoing SETI
experiments using this conservative approach will yield only
negative results.}

The picture sketched in the present study undermines the basic
tenets of the prevailing SETI philosophy. Outward migration of
advanced technological species should be taken into account in any
serious SETI project. Given the likely distances of an ATCs that
began migration tens of millions to billions of years ago
(Lineweaver 2001), they are not likely to know of our development.
While their observational capabilities probably allow them to
observe the Solar System, they are looking at it before
civilization developed. It is doubtful, to say the least, that
they would waste resources sending messages to planetary systems
possessing life, but quite uncertain (in light of the biological
contingency) to develop a technological civilization. Dolphins and
whales are quite intelligent and possibly even human-level
conscious (e.g.\ Browne 2004), but they do not have the ability to
detect signals from ATCs, and it is uncertain that they will ever
evolve such a capacity. By a mirror-image of such position, unless
one has concrete evidence of an ATC at a given locale, it would be
wasteful to direct SETI resources towards them. Ironically enough,
this can offer a rationale to some of the SETI sceptics, but
\textit{based on the entirely different overall astrobiological
picture\/} and with \textit{completely different practical
consequences}.

While fully recognizing that patience is a necessary element in
any search, we still wish to argue that the conventional SETI
(Tarter 2001; Duric and Field 2003, and references therein), as
exemplified by the historical OZMA Project, as well its later
counterparts (META, ARGUS, Phoenix, SERENDIP/ Southern SERENDIP,
etc.), notably those conveyed by NASA and the SETI Institute, is
fundamentally flawed. This is emphatically {\it not\/} due to the
real lack of targets, us being alone in the Galaxy, as
contact-pessimists in the mold of Tipler or Mayr have argued.
Quite the contrary, it is due to physical reasons underlying flaws
in the conventional SETI wisdom: in a sense the problem has
nothing to do with the universe itself, and everything to do with
our ignorance and prejudices. In this special sense, the flaws in
the currently prevailing views on SETI are less
excusable.\footnote{It is not just the present hypothesis which
leads to such a conclusion. Different views on the evolution of
ATCs, not based on the Intelligence Principle and the digital
perspective, lead to the same general idea. For example, this
applies to the ingenious idea that ATCs will transfer their
cognition into their environment (Karl Schroeder, private
communication), following recent studies on the distributed
natural cognition (e.g.\ Hutchins 1996). In these, as in other
suggested lines of "mainstream" ATC development, the approaches
currently favored by SETI projects will be fundamentally
misguided, i.e. ATCs remain undetectable by such approaches.}

Instead, much stronger emphasis on the ATC {\it manifestations and
traces\/} is the only serious recourse of practical SETI. Even if
they are not actively communicating with us, we could in principle
detect them and their astro-engineering activities. Their
detection signatures may be much older than their communication
signatures. Unless ATCs have taken great lengths to hide or
disguise their IR detection signatures, the terrestrial observers
should still be able to observe them at those wavelengths and
those should be distinguishable from normal stellar spectra. The
same applies to other un-natural effects, like the
antimatter-burning signatures (Harris 1986, 2002), or recognizable
transits of artificial objects (Arnold 2005). Search for
mega-projects such as Dyson shells, Jupiter Brains or stellar
engines are most likely to be successful in the entire spectrum of
SETI activities (Slysh 1985; Jugaku, Noguchi, and Nishimura 1995;
Jugaku and Nishimura 2003). Search for such astro-engineering
traces of ATCs should be primarily conducted in the infrared part
of the electromagnetic spectrum (Dyson 1960; Tilgner and
Heinrichsen 1998; Timofeev, Kardashev, and Promyslov 2000).
Ironically enough, surveys in the infrared have been proposed by
one of the pioneers of radio-astronomy, Nobel-prize winner Charles
H. Townes, although on somewhat different grounds (Townes 1983).
Bold and unconventional studies, such as Harris', Arnold's,
Slysh's, or the survey of Jugaku et al. and program proposed by
Tilgner and Heinrichsen (1998), represent still a minuscule
fraction of the overall SETI research. We dare suggest that there
is no real scientific reason for such situation: instead, it has
emerged due to excessive conservativeness, inertia of thought,
overawe of the "founding fathers", or some combination of the
three. The unconventional approach with emphasis on search for
ATCs' manifestations would loose nothing of the advantages of
conventional SETI before detection (e.g.\ Tough 1998), but the
gains could be enormous. In accordance with the motto of
Heraclitus, we should "expect the unexpected" if we desire genuine
SETI results; otherwise, we "won't find them" (i.e.\ traces of
ATCs).

The hypothesis presented here is falsifiable \textit{inter alia\/}
by extragalactic SETI observations. Extragalactic SETI has not
been considered very seriously so far (for notable exceptions see
Wesson 1990; Annis 1999b). The reason is, perhaps, the same old
comforting prejudice that we should expect specific (and,
conveniently, radio) signals. Since these are not likely
forthcoming over intergalactic distances (and two-way
communication desired by SETI pioneers is senseless here in
principle), there is no point in even thinking seriously about
extragalactic SETI. From the preceding, it is clear how
systematically fallacious such a view is: when we remove the cozy
assumption of specific SETI signals (together with the
second-order assumption of their radio nature), this view
collapses. On the contrary, extragalactic SETI would enable us to
probe enormously larger part of physical space as well as the
morphological space of possible evolutionary histories of ATCs.
(Of course, part of what we get ensemble-wise we loose time- and
resolution-wise.) In fact, the definition of Kardashev's Type III
civilization should prompt us to consider it more carefully, at
least for a sample of nearby galaxies, visible at epochs
significantly closer to us than the 1.8 Gyr difference between the
median age of terrestrial planets and the age of Earth
(All\`{e}gre et al.\ 1995; Lineweaver 2001). It could be argued
(although it is beyond the scope of the present study) that the
null result of extragalactic SETI observations so far (Annis
1999b) represents a strong argument against the viability of
Kardashev's Type III civilizations. While it remains a possibility
in the formal sense of being in agreement with the known laws of
physics, it seems that the type of pan-galactic civilization as
envisaged by Kardashev and other early SETI pioneers is either (i)
much more difficult to build (suggesting that the sample of $\sim
10^3$ normal spiral galaxies close enough and observed in high
enough detail is simply too small to detect even a single Type III
civilization), or (ii) simply not worth striving to. In contrast,
the concept of spatially smaller, compact, efficient ATCs
motivated by a convergent set of economic, ecological and/or
ethical premises, inhabiting fringes of the luminous matter
distribution presents to us more plausible alternative to the
conventional Type III picture. This will remain valid even if (for
some entirely different reason) the present hypothesis could not
account for Fermi's paradox in the Milky Way. The true test here
would be to detect signs of astro-engineering efforts at the
outskirts of nearby spiral galaxies (i.e.\ those which are seen at
about the same epoch as we are living in), and in their immediate
intergalactic vicinity. Observations of the edges of spiral
galaxies are notoriously difficult (e.g.\ Bland-Hawthorn, Freeman,
and Quinn 1997), but they are rapidly improving in both quality
and quantity. It is quite conceivable that they will give us the
first hint about the generic fate of advanced intelligent
communities.

\vspace{0.5cm}

\noindent {\bf Acknowledgements.} Invaluable technical help has
been received from Du\v san Indji\'c, Branislav K. Nikoli\'c, Sa\v
sa Nedeljkovi\'c, Vesna Milo\v sevi\'c-Zdjelar, Maja Bulatovi\'c,
Samir Salim, Nick Bostrom, Olga Latinovi\'c, Aleksandar Zorki\'c,
Milan Bogosavljevi\'c, Aleksandar B. Nedeljkovi\'c, and Vjera
Miovi\'c. Pleasant discussions with Zoran Kne\v zevi\'c, Irena
Dikli\'c, Ivana Dragi\'cevi\'c, Zoran \v Zivkovi\'c, Anders
Sandberg, Fred C. Adams, George Dvorsky, Petar V. Gruji\'c, Karl
Schroeder, Robin Hanson, Ma\v san Bogdanovski, James Hughes, Tanja
Beri\'c-Bjedov, Maja Jerini\'c, Mark A. Walker, Richard B.
Cathcart and Viorel Badescu have immensely contributed to
development of the ideas presented here. One of the authors (M. M.
\'C.) has been partially supported by the Ministry of Science and
Environmental Protection of Serbia through the project no.\ 1468,
"Structure, Kinematics and Dynamics of the Milky Way"; he also
uses the opportunity to thank the KoBSON Consortium of Serbian
libraries, which has enabled overcoming of the gap in obtaining
the scientific literature during the tragic 1990s in the Balkans.

\vspace{1.5cm}

\section*{References}

\refe Adams, F. C. and Laughlin, G. 1997, "A dying universe: the
long-term fate and evolution  of astrophysical objects," {\it
Rev.\ Mod.\ Phys.} {\bf 69}, 337-372.

\refe All\`{e}gre, C. J., Manh\`{e}s, G., and G\"{o}pel, C. 1995,
"The age of the Earth," {\it Geochim. Cosmochim. Acta\/} {\bf 59},
1445-1456.

\refe Andrews, D. G. 2003, "Interstellar Transportation using
Today's Physics," AIAA Paper 2003-4691, report to 39th Joint
Propulsion Conference \& Exhibit.

\refe Annis, J. 1999a, "An Astrophysical Explanation for the Great
Silence," {\it J. Brit. Interplan. Soc.} {\bf 52}, 19-22 (preprint
astro-ph/9901322).

\refe Annis, J. 1999b, "Placing a limit on star-fed Kardashev type
III civilisations," {\it J. Brit. Interplan. Soc.} {\bf 52},
33-36.

\refe Arnold, L. F. A. 2005, "Transit Lightcurve Signatures of
Artificial Objects," {\it Astrophys. J.}, in press (preprint
astro-ph/0503580).

\refe Asghari, N. et al. 2004, "Stability of terrestrial planets
in the habitable zone of Gl 777 A, HD 72659, Gl 614, 47 UMa and HD
4208," {\it Astron. Astrophys.} {\bf 426}, 353-365.

\refe Bada, J. L. 2004, "How life began on Earth: a status
report," {\it Earth Planet. Sci. Lett.\/} {\bf 226}, 1-15.

\refe Badescu, V. 1995, "On the radius of Dyson's sphere," {\it
Acta Astronautica\/} {\bf 36} 135-138.

\refe Badescu, V. and Cathcart, R. B. 2000, "Stellar engines for
Kardashev's type II civilizations," {\it J. Brit. Interplane.
Soc.} {\bf 53}, 297-306.

\refe Beaug\'e, C., Callegari, N., Ferraz-Mello, S., and
Michtchenko, T. A. 2005, "Resonance and stability of extra-solar
planetary systems," in \textit{Dynamics of Populations of
Planetary Systems}, Proceedings of the IAU Colloquium No. 197, ed.
by Z. Kneževic and A. Milani (Cambridge University Press,
Cambridge), 3-18.

\refe Binney, J. and Merrifield, M. 1998, {\it Galactic
Astronomy\/} (Princeton University Press, Princeton).

\refe Bland-Hawthorn, J., Freeman, K. C. and Quinn, P. J. 1997
"Where Do the Disks of Spiral Galaxies End?" {\it Astrophys. J.}
{\bf 490}, 143-155.

\refe Bostrom. N. 2000, "When Machines Outsmart Humans," {\it
Futures\/} {\bf 35}, 759-764.

\refe Bracewell, R. N. 1975, \textit{The Galactic Club:
Intelligent Life in Outer Space\/} (W. H. Freeman, San Francisco).

\refe Bradbury, R. J. 1997, "Jupiter \& Matrioshka Brains: History
\& References," preprint at {\tt
http://www.aeiveos.com/$\sim$bradbury/JupiterBrains/}.

\refe Bradbury, R. J. 2001, "Matrioshka brains," preprint at
\\ {\tt http://www.aeiveos.com/\\ $\sim$bradbury/MatrioshkaBrains/MatrioshkaBrains.html}

\refe Brillouin, L. 1962, {\it Science and Information Theory\/}
(Academic Press, New York).

\refe  Brin, G. D. 1983, "The great silence - the controversy
concerning extraterrestrial intelligent life," \textit{Q. Jl. R.
astr. Soc.} \textbf{24}, 283-309.

\refe Browne, D. 2004, "Do dolphins know their own minds?" {\it
Biology and Philosophy\/} {\bf 19}, 633-653.

\refe Carter, B. 1983, "The anthropic principle and its
implications for biological evolution,"  {\it Philos. Trans. R.
Soc. London A\/} {\bf 310}, 347-363.

\refe Cavicchioli, R. 2002, "Extremophiles and the Search for
Extraterrestrial Life," {\it Astrobiology\/} {\bf 2}, 281-292.

\refe Chernavskii, D. S. 2000, "The origin of life and thinking
from the viewpoint of modern physics," {\it Physics--Uspekhi\/}
{\bf 43}, 151-176.

\refe Clarke, J. N. 1981, "Extraterrestrial Intelligence and
Galactic Nuclear Activity," {\it Icarus\/} {\bf 46}, 94-96.

\refe Clube, S. V. M. 1978, "Does our galaxy have a violent
history?" \textit{Vistas in Astronomy\/} {\bf 22}, 77-118.

\refe Cohen, J. and Stewart, I. 2002, {\it What Does a Martian
Look Like?\/} (John Wiley \& Sons, Hoboken, New Jersey).

\refe Collar, J. I. 1996, "Biological Effects of Stellar Collapse
Neutrinos," {\it Phys. Rev. Lett.} {\bf 76},   999-1002.

\refe Cox, P., Kr\"{u}gel, E., and Mezger, P. G. 1986, "Principal
heating sources of dust in the galactic disk," {\it Astron.
Astrophys.} {\bf 155}, 380-396.

\refe Criswell, D. 1985, "Solar system industrialization:
Implications for interstellar migration," in \textit{Interstellar
Migration and the Human Experience}, ed. by B. Finney and E. Jones
(University of California Press, Berkeley), 50-87.

\refe \'Cirkovi\'c, M. M. 2004a, "Earths -- Rare in Time, Not
Space?" {\it J. Brit. Interplan. Soc.} {\bf 57}, 53-59.

\refe \'Cirkovi\'c, M. M. 2004b, "On the Temporal Aspect of the
Drake Equation and SETI," {\it Astrobiology\/} {\bf 4}, 225-231.

\refe \'Cirkovi\'c, M. M. 2004c, "Forecast for the Next Eon:
Applied Cosmology and the Long-Term Fate of Intelligent Beings,"
{\it Found. Phys.} {\bf 34}, 239-261.

\refe \'Cirkovi\'c, M. M. 2005, "'Permanence' -- An Adaptationist
Solution to Fermi's Paradox?" {\it J. Brit. Interplan. Soc.} {\bf
58}, 62-70.

\refe \'Cirkovi\'c, M. M. and Radujkov, M. 2001, "On the Maximal
Quantity of Processed Information in the Physical Eschatological
Context," {\it Serb. Astron. J.} {\bf 163}, 53-56.

\refe Dar, A. and De R\'{u}jula, A. 2002 "The threat to life from
Eta Carinae and gamma-ray   bursts," in \textit{Astrophysics and
Gamma Ray Physics in Space}, ed. by A. Morselli and P. Picozza
(Frascati Physics Series Vol. XXIV), pp. 513-523.

\refe Dick, S. J. 2003, "Cultural Evolution, the Postbiological
Universe and SETI," {\it Int. J. Astrobiology\/} {\bf 2}, 65-74.

\refe Drexler, K. E. 1992, "Molecular Manufacturing for Space
Systems: An Overview," {\it J. Brit. Interplan. Soc.} {\bf 45},
401-405.

\refe Dugi\'c, M. and \'Cirkovi\'c, M. M. 2002, "Quantum
Information Processing: The Case of Vanishing Interaction Energy,"
{\it Phys. Lett. A} {\bf 302}, 291-298.

\refe Duric, N. and Field, L. 2003, "On the Detectability of
Intelligent Civilizations in the Galaxy," \textit{Serb. Astron.
J.\/} \textbf{167}, 1-11.

\refe Dyson, F. J. 1960, "Search for Artificial Stellar Sources of
Infrared Radiation," {\it Science\/} {\bf 131}, 1667-1668.

\refe Dyson, F. J. 2003, "Looking for life in unlikely places:
reasons why planets may not be the best places to look for life,"
{\it Int. J.   Astrobiology\/} {\bf 2}, 103-110.

\refe Dyson, J. E. \& Williams, D. A. 1980, {\it Physics of the
interstellar medium\/} (Manchester University Press, Manchester).

\refe Ehrenfreund, P. et al. 2002, "Astrophysical and
astrochemical insights into the origin of life," {\it Rep. Prog.
Phys.} {\bf 65}, 1427-1487.

\refe Garriga, J., Mukhanov, V. F., Olum, K. D., and Vilenkin, A.
2000, "Eternal Inflation, Black Holes, and the Future of
Civilizations," {\it Int. J. Theor. Phys.} {\bf 39}, 1887-1900.

\refe Gonzalez, G., Brownlee, D., and Ward, P. 2001, "The Galactic
Habitable Zone: Galactic Chemical Evolution," {\it Icarus\/} {\bf
152}, 185-200.

\refe Gould, S. J. 1989, {\it Wonderful Life: The Burgess Shale
and the Nature of History\/} (W. W. Norton, New York).

\refe Hanson, R. 1998, "The great filter - are we almost past it?"
preprint available at \\ {\tt
http://hanson.gmu.edu/greatfilter.html} (1998).

\refe Harris, M. J. 1986, "On the detectability of antimatter
propulsion spacecraft," {\it Astrophys. Space Sci.} {\bf 123},
297-303.

\refe Harris, M. J. 2002, "Limits from CGRO/EGRET data on the use
of antimatter as a power source by extraterrestrial
civilizations," {\it J. Brit. Interplan. Soc.} {\bf 55}, 383-393.

\refe Hou, J. L., Prantzos, N., and Boissier, S. 2000, "Abundance
gradients and their evolution in the Milky Way disk," {\it Astron.
Astrophys.} {\bf 362}, 921-936.

\refe Hunt, G. E. 1978, "Possible climatic and biological impact
of nearby supernovae," {\it Nature\/} {\bf 271}, 430-431.

\refe Hutchins, E. 1996, \textit{Cognition in the Wild\/} (MIT
Press, Boston).

\refe Jugaku, J., Noguchi, K., and Nishimura, S. 1995, "A search
for Dyson spheres around late-type stars in the Solar
neighborhood," in \textit{Progress in the Search for
Extraterrestrial Life}, ed. by G. Seth Shostak (ASP Conference
Series, San Francisco), 381-185.

\refe Jugaku, J. and Nishimura, S. 2003, "A Search for Dyson
Spheres Around Late-type Stars in the Solar Neighborhood," in
\textit{Bioastronomy 2002: Life Among the Stars, Proceedings of
IAU Symposium \# 213}, ed. by R. Norris and F. Stootman (ASP
Conference Series, San Francisco), 437-438.

\refe Kardashev, N. S. 1964, "Transmission of information by
extraterrestrial civilizations," {\it Sov. Astron.} {\bf 8},
217-220.

\refe Knoll, A. H. and Bambach, R. K. 2000, "Directionality in the
history of life: diffusion from the left wall or repeated scaling
of the right?" in {\it Deep Time: Paleobiology's Perspective,} ed.
by D. H. Erwin and S. L. Wing (The Paleontological Society,
Lawrence, Kansas), 1-14.

\refe Korycansky, D. G., Laughlin, G., and Adams, F. C. 2001,
"Astronomical engineering: a strategy for modifying planetary
orbits," {\it Astrophys. Space Sci.} {\bf 275}, 349-366.

\refe Kuiper, T. B. H. and Brin, G. D. 1989, "Resource Letter
ETC-1: Extraterrestrial Civilization," {\it Am. J. Phys.} {\bf
57}, 12-18.

\refe Lachmann, M., Newman, M. E. J., and Moore, C. 2004, "The
physical limits of communication, or Why any sufficiently advanced
technology is indistinguishable from noise," {\it Am. J. Phys.}
{\bf 72}, 1290-1293.

\refe Lahav, N., Nir, S., and Elitzur, A. C. 2001, "The emergence
of life on Earth," {\it Progress in Biophysics \& Molecular
Biology\/} {\bf 75}, 75-120.

\refe Landauer, R. 1961, "Irreversibility and Heat Generation in
the Computing Process," \textit{IBM J. Res. Develop.} {\bf 5},
183-191.

\refe LaViolette, P. A. 1987, "Cosmic-ray volleys from the
Galactic center and their recent impacts on the Earth
environment," {\it Earth, Moon and Planets\/} {\bf 37}, 241-286.

\refe Lem, S. 1984, {\it His Master's Voice\/} (Harvest Books,
Fort Washington).

\refe Lem, S. 1987, {\it Fiasco\/} (Harcourt, New York).

\refe Lemarchand, G. A. 1995, "Detectability of Extraterrestrial
Technological Activities," \textit{SETIQuest\/} {\bf 1}, 3-13 (see
electronic version at \\ {\tt
http://www.coseti.org/lemarch1.htm}).

\refe Linde, A. D. 1990, {\it Inflation and Quantum Cosmology\/}
(Academic Press, San Diego).

\refe Linde, A. 1992, "Stochastic approach to tunneling and baby
universe formation," {\it Nuclear Physics B\/} {\bf 372}, 421-442.

\refe Lineweaver, C. H. 2001, "An Estimate of the Age Distribution
of Terrestrial Planets in the Universe: Quantifying Metallicity as
a Selection Effect," {\it Icarus\/} {\bf 151}, 307-313.

\refe Lineweaver, C. H. and Davis, T. M. 2002, "Does the Rapid
Appearance of  Life on Earth Suggest that Life Is Common in the
Universe?"  {\it Astrobiology\/} {\bf 2}, 293-304.

\refe Lineweaver, C. H., Fenner, Y., and Gibson, B. K. 2004, "The
Galactic Habitable Zone and the Age Distribution of Complex Life
in the Milky Way," {\it Science\/} {\bf 303}, 59-62.

\refe Livio, M. 1999, "How rare are extraterrestrial
civilizations, and when did they emerge?" {\it Astrophys. J.} {\bf
511}, 429-431.

\refe Lloyd, S. 2000, "Ultimate physical limits to computation,"
{\it Nature\/} {\bf 406}, 1047–1054.

\refe Mathis, J. S., Mezger, P. G., and Panagia, N. 1983,
"Interstellar radiation field and dust temperatures in the diffuse
interstellar matter and in giant molecular clouds," {\it Astron.
Astrophys.} {\bf 128} 212-229.

\refe McInnes, C. R. 2002, "Astronomical Engineering Revisited:
Planetary Orbit Modification Using Solar Radiation Pressure," {\it
Astrophys. Space Sci.} {\bf 282}, 765-772.

\refe McNamara, B. R., Nulsen, P. E. J., Wise, M. W., Rafferty, R.
A., Carilli, C., Sarazin, C. L., and Blanton, E. L. 2005, "The
heating of gas in a galaxy cluster by X-ray cavities and
large-scale shock fronts," {\it Nature\/} {\bf 433}, 45-47.

\refe Melott, A. L. et al. 2004, "Did a gamma-ray burst initiate
the late Ordovician mass extinction?" {\it Int. J. Astrobiol.}
{\bf 3}, 55-61.

\refe Mezger, P. G., Mathis, J. S., and Panagia, N. 1982, "The
Origin of the Diffuse Galactic Far Infrared and Sub-millimeter
Emission," {\it Astron. Astrophys.} {\bf 105}, 372-388.

\refe Minsky, M. 1973, in {\it Communication with Extraterrestrial
Intelligence (CETI)\/}, ed. by C. Sagan (MIT Press, Cambridge).

\refe Mojzsis, S. J., Arrhenius, G., McKeegan, K. D., Harrison, T.
M., Nutman, A. P., and Friend, C. R. L. 1996, "Evidence for life
on Earth before 3800 million years ago," {\it Nature\/} {\bf 384},
55-59.

\refe Moravec, H. P. 1988, \textit{Mind Children: The Future of
Robot and Human Intelligence\/} (Harvard University Press,
Cambridge).

\refe Noble, M., Musielak, Z. E., Cuntz, M. 2002, "Orbital
Stability of Terrestrial Planets inside the Habitable Zones of
Extrasolar Planetary Systems," {\it Astrophys. J.} {\bf 572},
1024-1030.

\refe Norris, R. P. 2000, "How old is ET?" in When SETI Succeeds:
The impact of   high-information Contact, ed. A. Tough (Foundation
for the Future,  Washington DC), pp. 103-105.

\refe Parkinson, B. 2005, "The carbon or silicon colonization of
the universe?" {\it J. Brit. Interplan. Soc.} {\bf 58}, 111-116.

\refe Phoenix, C. and Drexler, K. 2004, "Safe exponential
manufacturing," {\it Nanotechnology\/} {\bf 15}, 869-872.

\refe Prado, A. F. B. D. A. 1996, "Powered swingby,"
\textit{Journal of Guidance, Control, and Dynamics\/} {\bf 19},
1142-1147.

\refe Rose, C. and Wright, G. 2004, "Inscribed matter as an
energy-efficient means of communication with an extraterrestrial
civilization," {\it Nature\/} {\bf 431}, 47-49.

\refe Ruderman, M. A. 1974, "Possible consequences of nearby
supernova explosions for atmospheric ozone and terrestrial life,"
\textit{Science\/} {\bf 184}, 1079-1081.

\refe Sagan, C. 1975, "The recognition of extraterrestrial
intelligence," {\it Proc. R. Soc. Lond. B\/} {\bf 189}, 143-153.

\refe Sandberg, A. 1999, "The physics of information processing
superobjects: daily life among the Jupiter brains," {\it J. Evol.
Tech.} \textbf{5} \\ ({\tt http://www.jetpress.org/index.html}).

\refe Scalo, J. and Wheeler, J. C. 2002, "Astrophysical and
astrobiological implications of gamma-ray burst properties," {\it
Astrophys. J.} {\bf 566}, 723-737.

\refe Schaller, R. R. 1997, "Moore's law: past, present, and
future," {\it IEEE Spectrum}, June 1997, 53-59.

\refe Schroeder, K. 2002, {\it Permanence\/} (Tor Books, New
York).

\refe Slysh, V. I. 1985, "A search in the infrared to microwave
for astroengineering activity," in The Search  for
Extraterrestrial Life: Recent Developments, ed. by M. D.
Papagiannis (IAU, Reidel Publishing Co., Dordrecht), 315-319.

\refe Tadross, A. L. 2003, "Metallicity distribution on the
galactic disk," {\it New Ast.} {\bf 8}, 737-744.

\refe Tarter, J. 2001, "The Search for Extraterrestrial
Intelligence (SETI)," {\it Annu. Rev. Astron.  Astrophys.} {\bf
39}, 511-548.

\refe Thomas, B. C., Jackman, C. H., Melott, A. L., Laird, C. M.,
Stolarski, R. S., Gehrels, N., Cannizzo, J. K., and Hogan, D. P.
2005, "Terrestrial ozone depletion due to a Milky Way gamma-ray
burst," \textit{Astrophys. J.} {\bf 622}, L153-L156.

\refe Thorsett, S. E. 1995, "Terrestrial implications of
cosmological gamma-ray burst models," {\it Astrophys. J.} {\bf
444}, L53-L55.

\refe Tilgner, C. N. and Heinrichsen, I. 1998, "A program to
search for Dyson spheres with the infrared space observatory,"
\textit{Acta Astronautica\/} {\bf 42}, 607-612.

\refe Timofeev, M. Yu., Kardashev, N. S., and Promyslov, V. G.
2000, "A search of the IRAS database for evidence of Dyson
Spheres," \textit{Acta Astronautica\/} {\bf 46}, 655-659.

\refe  Tipler, F. J. 1986, "Cosmological limits on computation,"
{\it Int. J. Theor. Phys.} {\bf 25}, 617-661.

\refe Tough, A. 1998, "Positive consequences of SETI before
detection," {\it Acta Astronautica\/} {\bf 42}, 745-748.

\refe Townes, C. H. 1983, "At what wavelengths should we search
for signals from extraterrestrial intelligence?" {\it Proc. Natl.
Acad. Sci. USA\/} {\bf 80}, 1147-1151.

\refe Tucker, W. H. and Terry, K. D. 1968, "Cosmic Rays from
Nearby Supernovae: Biological Effects," \textit{Science\/} {\bf
160}, 1138-1139.

\refe Turnbull, M. C. and Tarter, J. C. 2003, "Target Selection
for SETI. I. A Catalog of Nearby Habitable Stellar Systems," {\it
Astrophys. J.} {\bf 145}, 181-198.

\refe Vinge, V. 1991, {\it A Fire upon the Deep\/} (Millenium,
London).

\refe Vulpetti, G. 1999, "On the viability of the interstellar
flight," {\it Acta Astronautica\/} {\bf 44}, 769-792.

\refe Ward, P. D. and Brownlee, D. 2000, \textit{Rare Earth: Why
Complex Life Is  Uncommon in the Universe\/} (Springer, New York).

\refe Ward, P. D. and Brownlee, D. 2002, \textit{The Life and
Death of Planet Earth: How the New Science of Astrobiology Charts
the Ultimate Fate of Our World\/} (Henry Holt and Company, New
York).

\refe Webb, S. 2002, {\it Where is Everybody? Fifty Solutions to
the Fermi's Paradox\/} (Copernicus, New York).

\refe Webber, W. R. 1987, "The interstellar cosmic ray spectrum
and energy density. Interplanetary cosmic ray gradients and a new
estimate of the boundary of the heliosphere," {\it Astron.
Astrophys.} {\bf 179}, 277-284.

\refe Weinberger, R. and Hartl, H. 2002, "A search for 'frozen
optical messages' from extraterrestrial civilizations," {\it Int.
J. Astrobiology\/} {\bf 1}, 71-73.

\refe Wesson, P. S. 1990, "Cosmology, Extraterrestrial
Intelligence, and a Resolution of the  Fermi-Hart Paradox," {\it
Q. Jl. R. astr. Soc.} {\bf 31}, 161-170.

\refe Wilson, P. A. 1994, "Carter on Anthropic Principle
Predictions," {\it Brit. J. Phil. Sci.} {\bf 45}, 241-253.

\refe Wright, E. L. et al. 1994, "Interpretation of the COBE FIRAS
CMBR Spectrum," {\it Astrophys. J.} {\bf 420}, 450-456.

\refe Zuckerman, B. 1985, "Stellar Evolution: Motivation for Mass
Interstellar Migrations," {\it Q. Jl. R. astr. Soc.} {\bf 26},
56-59.

\end{document}